# LIGHTWEIGHT DATASET FOR DECOY DEVELOPMENT TO IMPROVE IoT SECURITY

David Weissman[1]  and Anura P. Jayasumana[2]

[1]Department of Systems Engineering, Colorado State University, Fort Collins, CO, USA
[2]Department of Electrical and Computer Engineering, Colorado State University, Fort Collins, CO, USA

## ABSTRACT

*In this paper, the authors introduce a lightweight dataset to interpret IoT (Internet of Things) activity in preparation to create decoys by replicating known data traffic patterns. The dataset comprises different scenarios in a real network setting. This paper also surveys information related to other IoT datasets along with the characteristics that make our data valuable. Many of the datasets available are synthesized (simulated) or often address industrial applications, while the IoT dataset we present is based on likely smart home scenarios. Further, there are only a limited number of IoT datasets that contain both normal operation and attack scenarios. A discussion of the network configuration and the steps taken to prepare this dataset are presented as we prepare to create replicative patterns for decoy purposes. The dataset, which we refer to as IoT Flex Data, consists of four categories, namely, IoT benign idle, IoT benign active, IoT setup, and malicious (attack) traffic associating the IoT devices with the scenarios under consideration.*

## 1. INTRODUCTION

Internet of Things (IoT) remains one of the most vulnerable components of computer networking due to the vast number of expected interconnected and relatively low-cost devices. It is well documented that the need for cost-constrained IoT creates challenges for secure and privacy-safe products. Proprietary or non-standard offerings of IoT also add complexity since it is difficult to assess the security without wide access and open technical specifications for testing by experts. The closed nature of the design may inhibit the security community in guiding the vendor or announcing the vulnerabilities in the design and to propose stronger protection. Therefore, there is a need to evaluate security from proposed datasets that may be used to architect security programs.

In this paper, a measured dataset is introduced that offers a simple, yet comprehensive, approach to evaluate IoT data traffic for security purposes. The dataset is made available on an open-source platform. It provides network traffic events associated with the source and destination resources and the dataset comprises different scenarios using a reduced number of devices in a real network environment that may be typical for home use. The intention is to make available a quality dataset with a manageable number of packet events so that the analysis can be performed efficiently with limited computing resources. Unlike other datasets we examined, our IoT Flex

Data provides a reduced amount of traffic events in a realistic home IoT deployment setting, making it easier for researchers to assess with less computing resources.

Our proposed dataset captures IoT network traffic for 1-hour, 5-hour, and 10-hour uninterrupted intervals. As a result, the datasets are simpler to view in network diagnostic platforms, including Wireshark, Zui (formerly BRIM), and even excel spreadsheets, therefore avoiding frequently encountered constraints used when collecting and assessing data.

There are many IoT datasets that have been constructed and are available for viewing on open-source platforms. We provide an overview of references and surveys that mention or describe these alternatives. In general, datasets can be classified as being obtained in two general ways: 1) using real network traffic either from testbeds of IoT devices or historical recording of empirical network traffic in large-scale deployment or 2) synthesized data that is captured using computer simulation resources, imitating the device traffic. Our approach is categorized as a real network IoT dataset capture and the intention is to limit data events along with time duration so that deceptive characteristics of the environment can be implemented to improve security.

The value of any dataset, especially for cybersecurity objectives, is often based on certain parameters. Earlier works have demonstrated the importance of availability, quantity, and quality of the datasets. [1] It is worth noting that quantity of data, or the larger size of the dataset, may more likely contain representative samples. This may not necessarily suggest that a single large number of sequential events is preferred; instead recording more scenarios with limited events for each may be just as valuable. [2]

A discussion of the network configuration and the steps taken to prepare our dataset is presented. The dataset consists of four categories, namely, IoT benign idle, IoT benign active, IoT setup, and malicious (attack) traffic associating the IoT devices with the scenarios under consideration. For each scenario, except for IoT setup, three uninterrupted (1-hour, 5-hour, and 10-hour) datasets are recorded in PCAP (packet capture) format.

## 2. OBJECTIVE AND GOALS

The IoT Flex Dataset initiative is motivated by our desire to start small and expand the network with different devices and scenarios as our security research evolves. Initially, we were focused on invoking open-source datasets, but challenges with this approach became evident. This included the following:

1. Several datasets were too large to ingest in our collection tools. Even when datasets were compressed and downloaded, commonly used spreadsheet programs, such as excel have default limits of 65536 rows (events) or if increased may cause processing issues. Also, there are challenges with extracting datasets into PCAP viewing tools, such as Wireshark, without manipulation (trimming/splitting of data and other adjustments). We also observe that there were "labelled" datasets that were incompatible with viewing on platforms at the "flow" level (as opposed to packet level) unless certain parameters were superimposed, making the assessments more cumbersome.

2. Some datasets provided only limited descriptive information about the setup, devices deployed, and/or the scenarios in use. For instance, there were cases where Internet addressing changed during dataset captures, perhaps because of dynamic protocols in the network or other reasons which made continued assessment of the events difficult to interpret. In some of our attempts, we also had difficulty reaching authors or owners of the datasets to understand more about the scenarios.

3. There was difficulty with superimposing techniques, tactics, and procedures on original datasets to determine effects. In our case, this was perhaps the most conspicuous concern. Since access to the underlying network implementation can be restricted, the ability for adjustments (adding security solution algorithms) may be impeded. Augmentation of a countermeasure requires meticulous implementation and planning that further complicates the effort without control of the network.

These factors are not intended to negatively critique any dataset. In fact, we found all the datasets of value for their intended purpose. For our initiative, it became evident that a smaller, more contained dataset may have appeal. Also of importance is to make sure that the dataset is well documented, available widely, and easy to deploy. Hence our commitment to develop a dataset starting from scratch and keeping it lightweight for limited computing resources.

It is worth noting our goal using the proposed dataset is to assess security improvements. We evaluate scenarios of both benign (normal) traffic compared to conditions where IoT devices are under attack (malicious) traffic. Based on the comparison, IoT device characteristics, as portrayed with traffic patterns, can be reviewed and security solutions can be superimposed. Our ultimate objective is to impose a strategy that demonstrates improved security posture with reduced attack risk. Our goals are twofold:

1. Provide a simplified, lightweight set of data that can be interpreted with a high degree of significance and without excessive / heavy statistical processing.

2. Establish relationships between different scenarios of IoT device usage in a representative home environment.

Our expected differentiators include:

- IoT real devices (not virtualized)
- Manageable dataset (Shorter intervals and balanced events)
- Specific attack scenarios that can be easily identified
- Limited devices for easier analysis of addresses and communications
- Expandable - ability to overlay security measures for evaluation and result comparison

## 3. FLEX DATA COLLECTION

This section provides background on the data acquisition process. An overview is provided of our real network testbed and of the seven IoT devices that are deployed. The intention is to add replicated devices for security improvement measures as our research continues.

### 3.1. IoT Network Implementation

The IoT Flex Data is derived from a network of purchased IoT devices which are a smart camera, smart plugs, smart light bulbs, and Amazon Alexa devices. All are connected to the Internet for typical consumer use scenarios from the home network. Additionally, a virtual machine running on a hosted computer is configured to inject attack events on the IoT devices. Two classifications of traffic events contribute to the dataset. 1) traffic during normal operation and 2) traffic during an attack. Both normal and attack classes of events are recorded using Wireshark traffic monitoring off a switch with a mirror-port enabled attached to the wireless access point where the IoT devices are deployed.

When setting up the smart home IoT configuration, it became obvious that provisioning for all of the devices provided two options, either Bluetooth (a protocol for short-distance radio frequency) or IEEE 802.11 Wi-Fi connectivity. No other lower-layer network access option was available. In the implementation, Wi-Fi access is enabled with the radio frequency (RF) band determined by the capabilities of the IoT device, usually communicating over the 2.4GHz band. The local area segment containing the devices is covered using Ubiquiti Networks Access Point to extend Internet traffic from the interior router to the wireless network as illustrated in Figure 1.

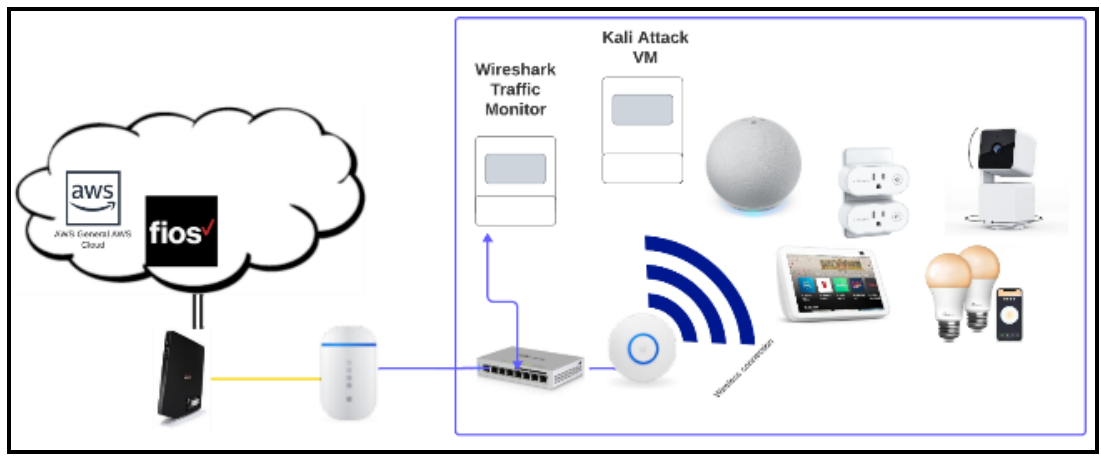

Figure 1. Network Implementation to Produce IoT Flex Data

To separate the IoT network environment from the other portions of the home network, Virtual Local Area Network (VLAN) tagging was used which enabled traffic monitoring for IoT devices of interest.

A span port from a switch capable of providing mirroring was positioned next to the Wi-Fi access point and captured network traffic in the VLAN segment which contained the IoT devices' communications. Note that attempting to capture IoT device traffic over the RF would have been more cumbersome, predominately due to Wi-Fi encryption and radio channel protocols, including frequency hopping. Monitoring of traffic in promiscuous mode was enabled on a packet diagnostic platform, Wireshark, to successfully capture events within the LAN segment. [3]

The network setup does not constrain use of other normal activity in the home. Figure 2 shows how the partitioned IoT network is isolated from other elements. It is worth noting that because the wireless access point is connected to a separate portion of the network, and our mirrored port is retrieving all traffic as seen by the wireless access point, the packet capture events (PCAP) contain ingress and egress traffic from the interface. Therefore, if viewing the traffic on a network monitoring tool, device communications from other superfluous connections may be observed, although we used an anonymization tool to mask sensitive identifiers just in case there was any vulnerability. The importance of keeping this traffic as part of the dataset is to make it appear as holistic and real as a network may be normally.

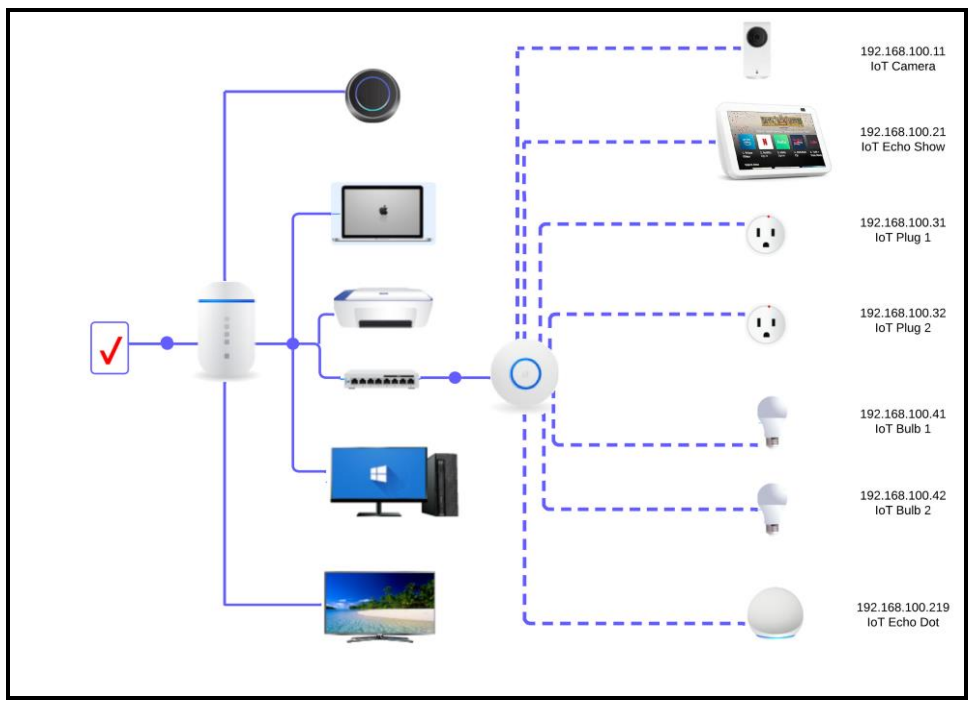

Figure 2. VLAN of IoT Devices in Wi-Fi Segment

## 3.2. IoT Devices and Computing

The deployment is based on four categories of IoT devices, ranging from a home security camera to lower traffic smart power plugs. A Wyze IoT high-fidelity video recording camera was selected with various controls to navigate the camera, control recordings, and/or listen (and observe) live or detected movements. Two Amazon Alexa Echo devices were deployed with one being used as a baseline, while the other is used to interact with offered services accordingly.

This paired solution was also implemented for Tuya IoT Lightbulbs and Tuya IoT plug outlet controllers. The network provisioning with our five applications, offers a simple way to interpret IoT devices and to understand how different types produce traffic patterns in isolation. We also observe how IoT devices interact with their peers or other types of IoT in the network. The implementation is simple, but expandable to other scenarios over time. Since the objective is to assess the smart home environment, the devices selected range from light traffic smart electrical plugs and controllable lightbulb to higher traffic home video cameras and Amazon Alexa interactive devices. These devices are connected on a virtual local area network with Wi-Fi. Table 1 highlights the baseline devices selected for our simplified implementation.

Table 1. Baseline deployment of devices

| Device | Vendor | Assigned IP Address | Hardware Details | Software Details |
|---|---|---|---|---|
| UAP-AC-Pro Wireless | Ubiquiti Networks | 192.168.1.101 | GbE Uplink PoE Wi-Fi 802.11a/b/g/n/ac 2.4 and 5 GHz | 802.1q VLAN |
| US-8-60W Port Switch | Ubiquiti Networks | 192.168.1.84 | 8 Port Switch (4 PoE) | Port Mirror Capable Non-Blocking |
| Unifi Dream Router | Ubiquiti Networks | 192.168.1.1 | 1 GbE WAN, 4x 1 GbE LAN Dual-core Arm Cortex A53 1.35 GHz Processor | |
| IoT Smart Camera Pan v3 | Wyze Labs, Inc. | 192.168.100.11 | 15/30 FPS , 1080p Full HD CPU 1.5 GHz Wi-Fi 802.11 b/g/n (2.4 GHz Wi-Fi only) | MQTT Publish/Subscribe H.264 video codec |
| Alexa Echo Show | Amazon Technologies Inc. | 192.168.100.21 | MediaTek MT 8163 Processor | MQTT Publish/Subscribe |
| Smart AC Electric Plug #1 | Tuya Smart Inc. | 192.168.100.31 | Wi-Fi 802.11 b/g/n 2.4 GHz only | MQTT Publish/Subscribe |
| Smart AC Electric Plug #2 | Tuya Smart Inc. | 192.168.100.32 | Wi-Fi 802.11 b/g/n 2.4 GHz only | MQTT Publish/Subscribe |
| Smart LightBulb #1 | Tuya Smart Inc. | 192.168.100.41 | A19 bulb, E26 Socket LED adjustable Wi-Fi 802.11 b/g/n, 2.4 GHz only | MQTT Publish/Subscribe |
| Smart LightBulb #2 | Tuya Smart Inc. | 192.168.100.42 | A19 bulb, E26 Socket LED adjustable Wi-Fi 802.11 b/g/n, 2.4 GHz only | MQTT Publish/Subscribe |
| Alexa Echo Dot | Amazon Technologies Inc. | 192.168.100.219 | MediaTek MT 8512 Processor | MQTT Publish/Subscribe |
| Desktop PC - DavidW-iMac | Apple Inc. | 192.168.100.232 | Intel Quad Core i5 3.2 GHZ 32 GB RAM | MacOS Monterey Kali Linux VMWare Fusion |
| Desktop - Network Monitor | Dell Inc. | 192.168.1.240 | Intel Core i7 1.80 GHz 64 GB RAM | Windows 11 Wireshark v4.2.3 TraceWrangler v0.6.7 |

Note the replication of a secondary device is made available for the IoT lightbulb and smart electrical plug. They are used for comparison purposes. Similar to a placebo, these added devices identify differences or any communications between the device's broker under evaluation and the secondary device.

### 3.3. Tools and Analytics

Several software platforms are part of IoT Flex Data production. They include the use of Wireshark, TraceWrangler, Kali Linux, and Zui/Zed (formerly BRIM) as summarized below:

**Wireshark** [3] is a network packet analyser that presents as much detail as possible. The platform summarizes captured data in layers, such as seeing each bit flow over a physical medium or observing framed data at higher layers. Wireshark is available as open source and is free.

**TraceWrangler** [4] is a network capture file toolkit that supports PCAP as well as the new PCAPng file format, which is now the standard file format used by Wireshark. The most prominent use case for TraceWrangler is the easy sanitization and anonymization of PCAP and PCAPng files. TraceWrangler is an open-source tool that we use to remove or replace sensitive data in the deployed network.

**Kali (NMAP, Hydra, HPING3)** [5] is an open-source, multi-platform distribution focused on performing Internet security tasks, such as penetration testing, security research, computer forensics, reverse engineering, and vulnerability management. We implement Kali as a virtual machine on our desktop Mac to generate the various attack scenarios on the IoT victim devices. Three software packages that are used, include NMAP for scanning, Hydra for bruit force of credentials and HPING3 for denial-of-service attack scenarios.

Computer Science & Information Technology (CS & IT) 133**Zui User Interface** [6] (formerly known as BRIM) provides a search query language enabling filtering and analytics from captured data traffic. Zui is used to analyse network traffic that was recorded by Wireshark in PCAP format which can then be examined at a connection (flow level) while also incorporating alerting as provided by intrusion detection systems, such as Suricata.

Table 2 summarizes the software and tools used for the assessments, although some of the platforms, including R-Studio and Excel, extend beyond generation of the dataset and are focused on the analytics.

Table 2. Software programs and platforms

| Purpose | Software | Description |
|---|---|---|
| Network Monitoring | Wireshark | Records network traffic in PCAP format |
| | Tracewrangler | Anonymize/mask local MAC addresses |
| | ZED/ZUI | Ingests PCAP and other formats. Compatibility with Zeek (Bro) Summarize/filter traffic at a connection layer |
| Security Alerts | Suricata | Intrusion Detection System Security alerting |
| Attack Engine | Kali Linux HPING3 NMAP Hydra | Denial of service and flooding Port and service scanning Brute force credentials |
| Statistics/Analysis | R-Studio | Versatile coding to summarize and graph results |
| | Excel | Summarize and tabularize results |

## 3.4. Data Capture Scenarios

This section highlights the generation of the traffic scenarios, both benign (normal) traffic and attack (abnormal) event traffic representing expected attacks in a real network environment. For the benign traffic production, we provide different time intervals and different activities. For attack scenarios, duration of recorded traffic will vary depending on the type of attack being used. A summary of the dataset scenarios made available are as follows and summarized in Table 3:

Normal Traffic Scenarios:

1. Startup activities for IoT lightbulb and IoT smart electrical plug. This activity involved a short capture interval of several minutes as IoT devices under examination were enrolled and powered-on in the IoT network segment to record traffic patterns.

2. All network IoT devices powered on, but without user-invoked activity (IoT idle/standby state). Here, IoT traffic from devices were recorded for different durations: one hour, five hours, and ten hours. The purpose was to achieve a baseline of events and how IoT device traffic may change between idle and activity. The devices in the network that were recorded include the IoT Smart Camera Pan V3, Amazon Echo Show, IoT smart electrical plugs #1 and #2, and IoT smart lightbulbs #1 and #2.

3. All network IoT devices powered on with IoT camera, Amazon Echo Show, IoT lightbulb #1, IoT electrical plug #1 performing tasks (IoT active state). Activities include turning the IoT lightbulb and IoT electrical plug on for exactly five minutes and then off for exactly five minutes for approximately 80% of each time-interval (one hour, five hours, and ten hours) of event traffic recorded in separate files. Similarly, user interaction was imposed on the IoT camera, such as listening to audio or observing video from the surveillance area, rotating video capture areas and setting up commands, such as activating movement sensors. For Alexa Echo Show, the user involvement would be commands to ask about news, play music and turn-off after numerous separate episodes of listening.

Attack Traffic Scenarios:

4. Reconnaissance: Scanning is often an early sign of malicious activity, especially if it is originating from an unknown resource. Using Kali Linux NMAP scanning, all devices on the network are identified including the ports (services) that may be open. Service scanning, often associated with identifying open ports and associated device addresses, may be an early indicator to take preventive action.

5. Bruit Force: Using Hydra credentials are injected and tracked to demonstrate the traffic characteristics associated with recurring username and password attempts.

6. Denial of Service: Using Hping3, adjustable conditions are configured to flood the IoT victim device. We attempt this on all devices separately to observe the overarching conclusion that may identify an attack is in process.

Table 3. Real traffic data for IoT devices

| Scenario | Duration | Devices/Victim | IP Address |
|---|---|---|---|
| **Setup** | | | |
| Power-On | | IoT Smart Electric Plug | 192.168.100.31 |
| | | IoT Lightbulb1 | 192.168.100.41 |
| **Benign** | | | |
| Idle State | | | |
| | 1 Hour | | |
| | 5 Hours | All IoT Devices | 192.168.100.0/24 |
| | 10 Hours | | |
| Active State | | 4 IoT Devices | 192.168.100.11 |
| | 1 Hour | IoT Camera, Alexa Echo Show, | 192.168.100.21 |
| | 5 Hours | IoT Smart Plug#1, IoT | 192.168.100.31 |
| | 10 Hours | Lightbulb#1) | 192.168.100.41 |
| **Attack** | | | |
| Denial of Service | | IoT Smart Electric Plug | 192.168.100.31 |
| Bruteforce | | IoT Camera Pan v3 | 192,168.100.11 |
| Reconaissance | | All IoT Devices | 192.168.100/24 |

Attack methods for IoT continue to expand. There are many different tactics that are associated with the hacker arsenal, and it remains difficult for organizations to stay current on defensive measures. Some of the more significant IoT attacks include loss of confidentiality, protocol and application integrity attacks, authentication attacks, denial of service, access control attacks, and attacks on physical security. As our research continues, additional attack scenarios may be added to the available dataset we post on Kaggle.

## 4. RELATED WORK

Based on our research, there are several surveys focused on IoT datasets for security. As a result, a review of academic publications and a search for publicly available IoT datasets was conducted. Searches of datasets were generated using web search engines and trade journal reviews, including IEEE Xplore, Google Scholar, along with data collection sites, such as SNAP (Stanford Large Network Dataset Collection), Kaggle, and GitHub. [7] [8] [9] [10] While there are many datasets available, several focus only on normal operation of IoT devices or, more frequently, on industrial sensors. Other datasets were performed in a simulated environment and not using real IoT deployments. Our findings also found that many datasets did not offer different scenarios, or a choice of duration.

Two important surveys of IoT datasets were reviewed as a starting point. *De Keersmaeker, Francois et al.* Compiled what may be perhaps the most inclusive survey with seventy-four datasets surveyed. The authors point out the diversity of data in the field of IoT analysis. The datasets in this survey are divided into several classifications and a summary of each of presented. [11] Similarly, Alex, Creado, et al, identified forty-four IoT datasets that were made available publicly. Several observations were made about limitations of existing datasets, which included the lack of documentation, less than realistic representation of IoT protocols and/or attack trends and lack of representative topologies. The author expressed a need to have a standard to discuss required attributes in dataset documentation. [12]

From the survey articles, we attempt to narrow down related datasets aligned to our IoT security research efforts. This focused listing in Table 4 attempts to notate representative attributes and security identifiers that we acknowledge as our IoT Flex Dataset was constructed.

Table 4. Listing of comparable datasets related to IoT Security

| Dataset | Originator | Affiliatiion | Year |
|---|---|---|---|
| DARPA Off-Line Intrusion Detection | Lippmann *et al.* [13] | DARPA (Defense Advanced Research Projects Agency) | 1998/1999 |
| Bot-IoT Dataset | Koroniotis *et al.* [14] | UNSW (University of New South Wales) | 2018 |
| UNSW-NB15 | Moustafa et al. [15] [16] | UNSW (University of New South Wales) | 2019 |
| CIC-IOT | Dadkhah *et al.* [17] | CIC (Canadian Institute for Cybersecurity) | 2022 |
| Aposemat IoT-23 | Garcia *et al.* [18] | CVUT (Czech Technical University) | 2020 |
| AIoT-Sol Dataset | Min *et al.* [19] | Mahidol University, Thailand | 2018 |

**1998/99 DARPA off-line intrusion detection** [13]

Among the earlier datasets identified was the 1999 DARPA work on intrusion detection focusing on computer networking in general. The dataset was a real-world evaluation of computing devices and captures various attack type traffic, including denial-of-service, port scanning and other malicious intrusions. The dataset was crafted on a wired network infrastructure and had no focus on the emergence of IoT as this ecosystem evolved much later.

**IoT Botnet** [14]

The Bot-IoT dataset includes normal IoT network traffic along with a variety of attacks. The dataset contains DDoS, data exfiltration, and service (port) scans, similar to our dataset. This dataset is a very popular reference and is used frequently for simulated machine learning objectives. However, the IoT endpoints are simulated so actual smart home devices were not a part of the data that would be deployed in the consumer market. We also had limited ability to add prevention mechanisms to determine how the attack scenarios can be reduced with an overlay security approach.

**UNSW-NB15** [15] [16]

UNSW-NB15 was initially released in 2015. This dataset is comprised of normal and abnormal (attack) traffic events. Two servers were configured to distribute the normal network traffic and a third server was deployed to generate attack traffic. Both packet-based and flow-based features were extracted from the raw network packets using analytic platforms Argus and Bro (now called Zeek). Packet-based features are extracted from the packet header and its payload. In contrast, flow-based features are generated using the sequencing of packets, from a source to a destination, traveling in the network. In our IoT Flex Data, we use similar flow-based features to capture the characteristics of IoT devices under both benign and attack scenarios. We use real IoT devices in our configuration.

**CIC IoT Dataset** [17]

The Canadian Institute for Cybersecurity sponsored a project to generate IoT datasets profiling different IoT devices with different protocols, including IEEE 802.11 (Wi-Fi), Zigbee and Z-Wave. The objectives were to configure various IoT devices and analyse the behaviour exhibited. The goal of the dataset was intended to analyse network traffic when devices were idle and then powered on. Furthermore, the dataset captured network traffic of devices in active benign conditions and then under an attack scenario. For this work, 40 IoT devices were used, and data was captured in a simulated environment for five days. The generation of CIC-IDS2017 [9] was done in an emulated environment for 5 days.

In comparison, our IoT Flex Data uses only seven IoT devices and curtails the duration to under ten hours. While CIC IoT data is a very popular source for a dataset, our intention was to simplify the traffic load under examination.

**IoT-23** [18]

IoT-23 was developed in the Stratosphere Laboratory at CTU University in the Czech Republic. It is one of the more recently released datasets of network traffic from Internet of Things (IoT) devices. It has 20 malware captures executed in IoT devices, and 3 captures for benign IoT devices traffic. It was first published in January 2020, with captures ranging from 2018 to 2019. Similar to our proposed work, IoT-23 data provides benign scenarios for a Philips HUE smart LED lamp, an Amazon Echo home intelligent personal assistant and a Somfy smart door lock. Also similar, the three IoT devices are real hardware and not simulated. In our work, we had difficulty superimposing defensive security measures that could be evaluated on this dataset. Instead, we preferred to generate results from a running network that would be easier to set up or replicate as opposed to generating specialized coding to adapt to this dataset that was available.

**AIoT-Sol Dataset** [19]

IoT-Sol dataset uses a real device network to incorporate attack techniques in the IoT datasets. A total of 17 attack types are selected, such as network attacks, web attacks, web IoT message protocol attacks, it also includes the typical attack types available in most IoT datasets, such as denial-of-service (DoS) attacks. It also contains a simulation of realistic IoT network traffic in a normal operation scenario.

## 5. PRELIMINARY REPRESENTATIVE FINDINGS

As we begin to make statistical inferences of our data, there were several relevant observations that became evident. We want to convey two of these findings and follow up later with more comprehensive details in another publication.

First, it was observed that even when IoT devices were in a non-active state IoT devices were transmitting and receiving data outside our network environment. It was expected that in an "idle" state there would be no traffic as no action was being taken by users of the devices. However, this was not the case; in fact, upon power-on of the IoT devices, traffic was noticeable to Amazon Web Services (AWS), perhaps because the cloud services that control/broker the devices which are then aggregated there. Upon further assessment, it became clear that the AWS environment contains the MQTT (Message Queueing Telemetry Transport). Second, in the case of IoT lightbulbs and IoT power outlet devices, lateral communications occurred between active and idle devices; this was not initially expected either.

With respect to attack-generated scenarios, data statistics clearly indicated interruption. Preliminary data shown below in Figure 3 depicts an IP address of the device where a Denial of Service (DOS) attack was performed using Kali Linux. Specifically, the transmit and receive packet statistics show IoT Plug 1 device (192.168.100.31) unable to transmit due to traffic flooding.

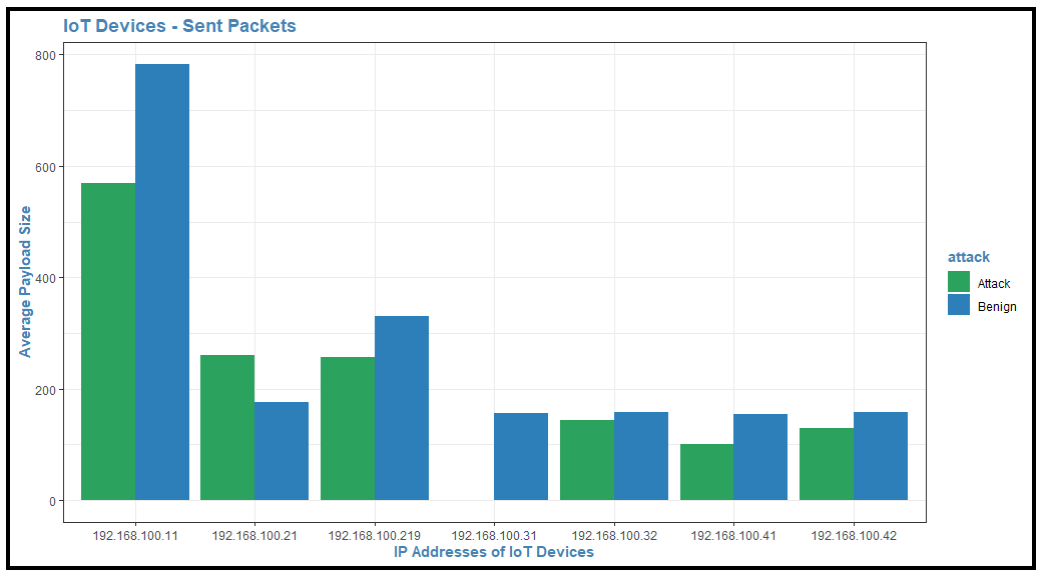

Figure 3. Representative Results From a Denial of Service Attack on IoT Plug 1 Device

## 6. FUTURE WORK

The proposed dataset is primarily motivated to continue future research around improving security for IoT networks. We plan to overlay defensive mechanisms and observe the responsiveness of the devices.

In the future, our plan is to expand the dataset offering to include traffic when defensive measures are invoked to counteract the attack scenarios. Additionally, we intend to provide analytical results that would demonstrate an improved secure network posture with the selected defensive strategies.

Our work in general focuses on deception techniques with the objective of entangling intruders as opposed to attracting them once in the network.[20] In other words, implementations are deployed with proportional decoys throughout the environment so that the probabilistic outcome would be extremely low for the attacker to have succeeded in reaching authentic IoT endpoints altogether. Solutions in this regard have progressed to include enriched deception by being widely distributed and being capable of stealthy interactions with a better determination of the attacker's intent and approach.

As an attacker takes action and attempts to make lateral moves to other endpoints, alerts are sent, and deception is activated. At this point, decoys can be multiples of the real system resulting in a perceived larger number of devices than in use. The attacker's attempt is directed to a trap server instead of the real assets that were targeted. With numerous deceptions invoked, the likelihood is significant that the intruder makes an incorrect decision and does not realize it. This compels the attacker to perform more work and leave more trails, positive characteristics for an investigation aligned with a defensive strategy.

The deceptive elements are typically maintained in the virtual server(s) that spread the decoys across each family class of protection defined. The information attained from real-time forensics of the trap logs can be channelled to the management console for countermeasures. Figure 4 depicts a generic representation of a deception solution as it relates to IoT.

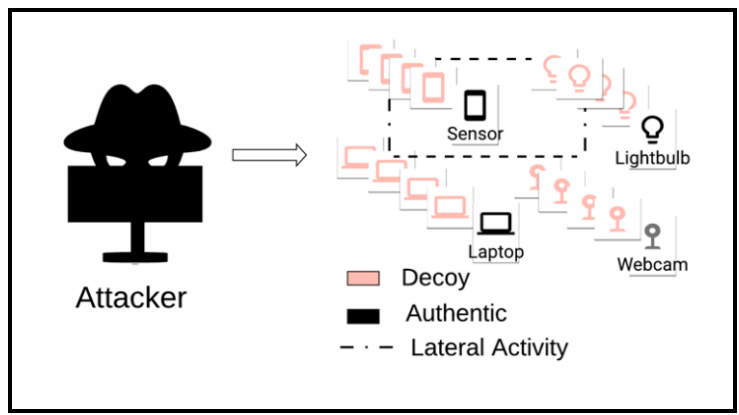

Figure 4. Deceptive View from Attacker Perspective

## 7. CONCLUSIONS

The objective of this discussion was to present and make available a realistic IoT network in a home environment that would be able to produce a lightweight dataset for use in academia, industry and for other purposes. An attempt was made to be descriptive, yet concise, on the details of the IoT devices under consideration, the network configuration, and the software tools being used to investigate responsiveness under different scenarios. Network traffic of representative smart home IoT devices was recorded. Our intention was to accurately describe the configuration and make the investigative durations manageable so it could be replicated and evaluated by others. Our main contribution is to add perspectives from this data that may not have been entirely captured in the past and that may be easier to analyse than much larger datasets from the past.

AUTHORS

**David Weissman** is a Ph.D. candidate at Colorado State University in the Systems Engineering Department. His current area of research is in cybersecurity solutions for government and business infrastructure advancement. He is currently focused on Internet of Things security, particularly relating the use of deception technology to protect and reduce risk to critical assets. He has worked in advanced technical, business, and financial professional capacities for 25+ years involving both government and commercial sectors.

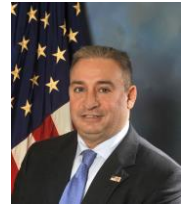

**Anura P.** Jayasumana is a Professor in the Electrical & Computer Engineering at Colorado State University where he holds a joint appointment in Computer Science. He is the Director of the Information Science and Technology Center (ISTeC) at CSU, a university wide organization for promoting research, teaching and service in information sciences and technologies. He received a Ph.D. and M.S. in Electrical Engineering from Michigan State University and B.Sc. in Electronic and Telecommunications Engineering with First Class Honors from University of Moratuwa, Sri Lanka. His current research interests include mining of network-based data for radicalization detection, machine learning techniques for graphs, Internet of Things, detection of weak distributed patterns, and synthetic data generation. His research has been funded by DARPA, NSF, DoJ/NIJ, and industry. He served as a Distinguished Lecturer of the IEEE Communications Society (2014-17) and is currently ACM Distinguished Speaker. He has served extensively as a consultant to numerous companies ranging from startups to Fortune 100 companies, and is a member of Phi Kappa Phi, ACM and IEEE.

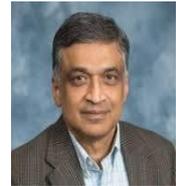